\documentclass[%
 reprint,
 superscriptaddress,
 showpacs,preprintnumbers,
 amsmath,amssymb,
 aps,
 prb,
]{revtex4-1}

\usepackage{graphicx}% Include figure files
\usepackage{dcolumn}% Align table columns on decimal point
\usepackage{bm}% bold math
\usepackage{hyperref}% add hypertext capabilities
\usepackage{color}

\begin{document}

\preprint{}

% Force line breaks with \\
\title{Phonon softening in paramagnetic bcc Fe and relationship with pressure-induced phase transition}

\author{Yuji Ikeda}
\email{ikeda.yuji.6m@kyoto-u.ac.jp}
\affiliation{Center for Elements Strategy Initiative for Structure Materials (ESISM), Kyoto University, Kyoto 606-8501, Japan}
\author{Atsuto Seko}
\affiliation{Center for Elements Strategy Initiative for Structure Materials (ESISM), Kyoto University, Kyoto 606-8501, Japan}
\affiliation{Department of Materials Science and Engineering, Kyoto University, Kyoto 606-8501, Japan}
\author{Atsushi Togo}
\affiliation{Center for Elements Strategy Initiative for Structure Materials (ESISM), Kyoto University, Kyoto 606-8501, Japan}
\author{Isao Tanaka}
\email{tanaka@cms.mtl.kyoto-u.ac.jp}
\affiliation{Center for Elements Strategy Initiative for Structure Materials (ESISM), Kyoto University, Kyoto 606-8501, Japan}
\affiliation{Department of Materials Science and Engineering, Kyoto University, Kyoto 606-8501, Japan}
\affiliation{Nanostructures Research Laboratory, Japan Fine Ceramics Center, Nagoya 456-8587, Japan}

%\affiliation{
%\affiliation{
%\affiliation{%

\date{\today}

%%%%%%%%%%%%%%%%%%%%%%%%%%%%%%%%%%%%%%%%
\begin{abstract}
Structural stability of paramagnetic (PM) body-centered cubic (bcc) Fe
under pressure is investigated based on first-principles phonon calculations.
Spin configurations of the PM phase are approximated
using a binary special quasi-random structure (SQS) with a supercell approach.
The behavior of phonon modes can be associated with pressure-induced
phase transitions to the face-centered
cubic (fcc) and hexagonal close-packed (hcp) structures as follows:
For the PM phase, it is found that
the low-frequency transverse mode at the N point (N$_4^-$ mode),
which corresponds to a bcc-hcp phase transition pathway,
exhibits strong softening under isotropic volume compression.
The frequency of this mode becomes zero by $2\%$ volume decrease
within the harmonic approximation.
This result is not consistent with the experimental fact that
phase transition
from the PM bcc to hcp phases does not occur under volume compression.
The seeming contradiction can be explained
only when anharmonic behavior of the N$_4^-$
mode is taken into consideration;
a potential energy curve along the N$_4^-$ mode
becomes closer to a double-well shape for the PM phase
under the volume compression.
On the other hand,
softening of the longitudinal mode at the 2/3[111] point
under the volume compression
is also found for the PM phase,
which indicates the pressure-induced bcc-fcc phase
transition along this mode.
Such behaviors are not seen in ferromagnetic (FM) bcc Fe,
implying that the magnetic structure plays essential roles
on the phase transition mechanism.
\end{abstract}
%%%%%%%%%%%%%%%%%%%%%%%%%%%%%%%%%%%%%%%%

\pacs{71.15.Mb, 75.50.Bb, 63.20.dk}% PACS, the Physics and Astronomy
\maketitle

%\tableofcontents

%%%%%%%%%%%%%%%%%%%%%%%%%%%%%%%%%%%%%%%%
\section{INTRODUCTION}
\label{sec:introduction}
%%%%%%%%%%%%%%%%%%%%%%%%%%%%%%%%%%%%%%%%

\begin{figure}[tbp]
\begin{center}
\includegraphics[width=\linewidth]{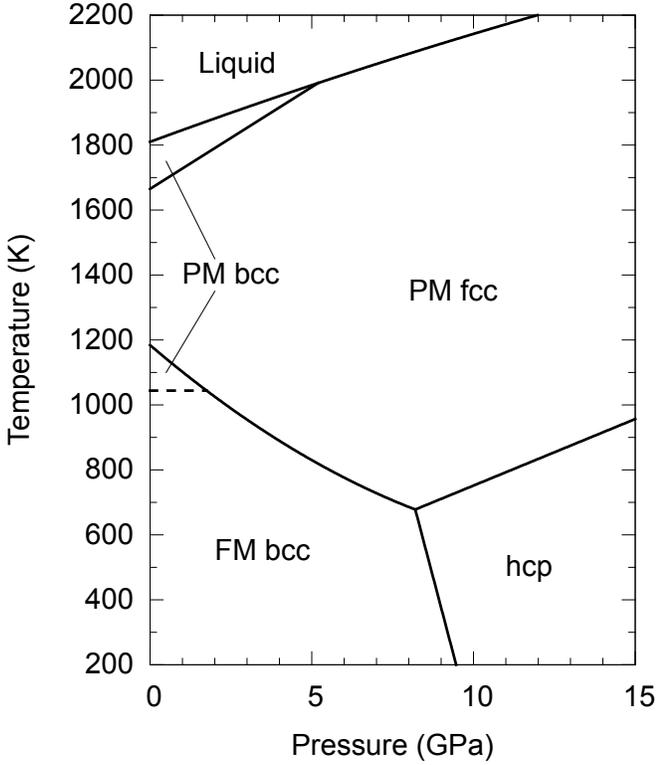} % phase_diagram
\end{center}
\caption{
%(Color online)
Experimental phase diagram of Fe.
\cite{Leger1972,Strong1973,Shen1998,Klotz2008}
Solid and dashed bold lines represent structural and magnetic phase boundaries,
respectively.
For the magnetic structure of the hcp phase,
both nonmagnetic (NM) and anti-ferromagnetic phases have been proposed.
\label{fig:phase_diagram}
}
\end{figure}

The elemental Fe is widely known to exhibit several structural and magnetic
phases, which have been investigated over many years.
Figure~\ref{fig:phase_diagram} shows an experimental phase diagram
for the elemental Fe.
\cite{Leger1972,Strong1973,Shen1998,Klotz2008}
Three stable phases, namely body-centered cubic (bcc),
face-centered cubic (fcc), and hexagonal close-packed (hcp) exist
as solid phases.
They can be further classified by magnetic structures, i.e.,
ferromagnetic (FM) and paramagnetic (PM) phases.
At ambient pressure, the stable phase changes as
FM bcc $\rightarrow$ PM bcc $\rightarrow$ PM fcc $\rightarrow$ PM bcc
with increase of temperature.
At room temperature, the elemental Fe undergoes a pressure-induced bcc-hcp
phase transition at 9.2~GPa.
Above the temperature of the triple point where the three phases coexist
at 678~K and 8.2~GPa,
bcc-fcc phase transition occurs.
%The complicated phase transition behaviors are closely related to the magnetic
%structures.
%

Analysis of structural stability is an important approach
to investigate the phase transition behavior.
From the theoretical viewpoint,
the structural stability of crystalline materials
can be analyzed using a combination of density functional theory (DFT)
and phonon calculations.
Therefore, an additional method is required to analyze the structural stability
of high temperature phases.
The PM phase of Fe is a typical example.
%The structural stability at high temperature is, however,
%sometimes difficult to investigate by the DFT-based approach,
%since the DFT is the theory at the absolute zero temperature in principle.
%This problem is very significant for magnetic crystalline materials
%such as Fe, which becomes the PM phase at sufficiently high temperature. 
%
In recent years,
several calculation methods for obtaining phonon frequencies of
the PM phase have been proposed.
Leonov \textit{et al.} reported a method based on
the dynamical mean-field theory (DMFT).
\cite{Leonov2012}
They succeeded to reproduce experimental phonon dispersion relations
for both PM bcc and PM fcc Fe.
\cite{Zarestky1987, Neuhaus1997}
K\"{o}rmann \textit{et al.} developed a method for calculating phonon
frequencies for the PM phase
\cite{Kormann2012}
based on the binary special quasi-random structure (SQS).
\cite{Zunger1990}
In this method, force constants for the PM phase were obtained by
a symmetrization process.
Their results for the PM bcc and PM fcc Fe were also consistent with the
experimental ones.
\cite{Zarestky1987, Neuhaus1997}
Ruban and Razumovskiy proposed another way to approximate the PM phase
using several spin spiral states.
\cite{Ruban2012}

The present study aims at elucidating the behavior of phonons in the PM phase
under pressure by means of systematic DFT-based phonon calculations.
We adopt a method equivalent to that reported in Ref.~\citenum{Kormann2012}
to obtain phonon frequencies of the PM phase.
Volume dependence of the phonon frequencies and the potential energy curves
along phonon modes are investigated
in order to examine their anharmonic behaviors.
Results are mainly discussed from the viewpoint of the pressure-induced
phase transition for the PM bcc Fe.

%%%%%%%%%%%%%%%%%%%%%%%%%%%%%%%%%%%%%%%%
\section{METHODOLOGY}
%%%%%%%%%%%%%%%%%%%%%%%%%%%%%%%%%%%%%%%%

%%%%%%%%%%%%%%%%%%%%%%%%%%%%%%%%%%%%%%%%
\subsection{Modeling of PM phase}
%%%%%%%%%%%%%%%%%%%%%%%%%%%%%%%%%%%%%%%%

In principle, a magnetic state at finite temperature is described as the
statistical average of all possible spin configurations.
For the PM phase, instead of evaluating the statistical average,
a completely disordered spin configuration is often used for simplicity.
The SQS mimics such a disordered configuration within a periodic structure.
\cite{Zunger1990} % A. Zunger Phys. Rev. Lett. 65 353 (1990)
The SQS originates from the cluster expansion method which corresponds to
generalized Ising model.
Here the binary SQS is applied to the completely disordered collinear spin configuration.
The SQS is generated from a supercell obtained by
$2 \times 2 \times 2$
isotropic expansion of the conventional bcc unit cell.
The isotropic expansion is adopted here because force constants of the PM phase
are easily computed.
The SQS generated from the
$2 \times 2 \times 2$
supercell was reported to be sufficient to reproduce the experimental phonon
dispersion relations of the PM bcc Fe.\cite{Kormann2012}
With the constraint of the supercell, pair correlation functions up to
several nearest neighbors are optimized using simulated annealing procedure
as implemented in the CLUPAN code.
\cite{
Seko2006, % PhysRevB.73.094116 % A. Seko, (2006)
Seko2009} % PhysRevB.80.165122 % A. Seko, (2009)

%%%%%%%%%%%%%%%%%%%%%%%%%%%%%%%%%%%%%%%%
\subsection{Phonon calculations}
%%%%%%%%%%%%%%%%%%%%%%%%%%%%%%%%%%%%%%%%

Phonon frequencies are calculated by means of the finite-difference method
with a displacement of 0.01~\AA.
For the FM and NM phases,
atomic displacements are given to the $2 \times 2 \times 2$ supercell of a
conventional bcc unit cell, and the forces acting on atoms are collected.
Force constants are calculated from the set of forces
in the least-squares sense.
\cite{Parlinski1997,Chaput2011}

Before the depiction of how to obtain the force constants of the PM phase,
we briefly describe calculation procedure of the force constants
for the FM and NM phases.
%in order to explain the calculation procedure of the PM state in a relatively
%simple way later in this subsection.
%For more details on the calculation procedure, see APPENDIX in Ref.~\citenum{Chaput2011}.
We use $R$ and $\tau$ for the index of unit cells and the index of atoms
in the corresponding unit cell, respectively.
A second-order force constant is denoted as
$\Phi^{\alpha\beta}_{R_{i}\tau_{i},R_{j}\tau_{j}}$,
where the superscripts $\alpha$ and $\beta$ are used for the indices of
Cartesian coordinates,
and the subscripts $R_{i}\tau_{i}$ and $R_{j}\tau_{j}$ specify a pair of atoms.
For computational convenience,
the force constants for a pair of atoms, $R_{1}\tau_{1}$ and $R_{2}\tau_{2}$,
are represented by a $9 \times 1$ matrix $\mathbf{P} (R_1\tau_1, R_2\tau_2)$
given by
\begin{align}
\mathbf{P}(R_{1}\tau_{1}, R_{2}\tau_{2})
=
\begin{pmatrix}
\Phi^{xx}_{R_{1}\tau_{1}, R_{2}\tau_{2}} \\
\Phi^{xy}_{R_{1}\tau_{1}, R_{2}\tau_{2}} \\
\Phi^{xz}_{R_{1}\tau_{1}, R_{2}\tau_{2}} \\
\Phi^{yx}_{R_{1}\tau_{1}, R_{2}\tau_{2}} \\
\Phi^{yy}_{R_{1}\tau_{1}, R_{2}\tau_{2}} \\
\Phi^{yz}_{R_{1}\tau_{1}, R_{2}\tau_{2}} \\
\Phi^{zx}_{R_{1}\tau_{1}, R_{2}\tau_{2}} \\
\Phi^{zy}_{R_{1}\tau_{1}, R_{2}\tau_{2}} \\
\Phi^{zz}_{R_{1}\tau_{1}, R_{2}\tau_{2}}
\end{pmatrix}.
\end{align}
In addition, an atomic displacement applied to the atom $R_{1}\tau_{1}$
is described as a $3 \times 9$ matrix $\mathbf{U}_i (R_1\tau_1)$ given by
\begin{align}
\mathbf{U}_i(R_{1}\tau_{1})
&=
\begin{pmatrix}
1 & 0 & 0 \\
0 & 1 & 0 \\
0 & 0 & 1
\end{pmatrix}
\otimes
\begin{pmatrix}
\Delta x &
\Delta y &
\Delta z
\end{pmatrix},
%&=
%\begin{pmatrix}
%\Delta x & \Delta y & \Delta z & 0 & 0 & 0 & 0 & 0 & 0 \\
%0 & 0 & 0 & \Delta x & \Delta y & \Delta z & 0 & 0 & 0 \\
%0 & 0 & 0 & 0 & 0 & 0 & \Delta x & \Delta y & \Delta z
%\end{pmatrix},
\end{align}
%where $\Delta \mathbf{r}_{R_{2}\tau_{2}}
%=
%(\Delta r^{x}_{R_{2}\tau_{2}}
% \Delta r^{y}_{R_{2}\tau_{2}}
% \Delta r^{z}_{R_{2}\tau_{2}})$
%represent the displacement given to the atom specified by $R_{2}\tau_{2}$.
where
$\Delta {x}$,
$\Delta {y}$, and
$\Delta {z}$
represent Cartesian components of the displacement given to the atom $R_{1}\tau_{1}$,
and the subscript $i$ is for the index of atomic displacements.
For the system with the displacement $\mathbf{U}_i (R_1\tau_1)$,
a force acting on the atom $R_2\tau_2$
can be written in the form of a $3 \times 1$ matrix $\mathbf{F}_i (R_2\tau_2)$ 
as
\begin{align}
\mathbf{F}_i(R_2\tau_2)
=
\begin{pmatrix}
F_x \\
F_y \\
F_z
\end{pmatrix},
\end{align}
where $F_x$, $F_y$, and $F_z$ are Cartesian components of the force
acting on the atom $R_2\tau_2$.
We can obtain $\mathbf{F}_i (R_{2}\tau_{2})$ using
$\mathbf{P} (R_{1}\tau_{1}, R_{2}\tau_{2})$ and $\mathbf{U}_i (R_{1}\tau_{1})$ as
\begin{align}
\mathbf{F}_i(R_{2}\tau_{2})
=
-
\mathbf{U}_i(R_{1}\tau_{1})
\mathbf{P}(R_{1}\tau_{1}, R_{2}\tau_{2}).
\end{align}
Simultaneous equations of different atomic displacements for the pair of atoms
are then combined as 
\begin{align}
\begin{pmatrix}
\mathbf{F}_{1} \\
\mathbf{F}_{2} \\
\vdots
\end{pmatrix}
=
-
\begin{pmatrix}
\mathbf{U}_{1} \\
\mathbf{U}_{2} \\
\vdots
\end{pmatrix}
\mathbf{P}.
\label{eq:forces}
\end{align}
With a sufficient number of atomic displacements,
Eq.~(\ref{eq:forces}) can be solved by the pseudoinverse such as
\begin{align}
\mathbf{P}
=
-
\begin{pmatrix}
\mathbf{U}_{1} \\
\mathbf{U}_{2} \\
\vdots
\end{pmatrix}^{+}
\begin{pmatrix}
\mathbf{F}_{1} \\
\mathbf{F}_{2} \\
\vdots
\end{pmatrix}.
\label{eq:force_constants}
\end{align}
To recover site symmetry around the displaced atom,
we apply the site-symmetry operations to the atomic system and obtain
new displacements and corresponding forces.
By adding these displacements and forces to the simultaneous
equations shown in Eq.~(\ref{eq:forces}),
we can obtain symmetry-recovered force constants.
Note that for the atoms at the crystallographically equivalent sites,
we need to apply displacements only to one of them;
by considering the symmetry of the structure,
force constants for the other atoms can be derived without calculations via
Eq.~(\ref{eq:force_constants}).

\begin{figure}[tbp]
\begin{center}
\includegraphics[width=\linewidth]{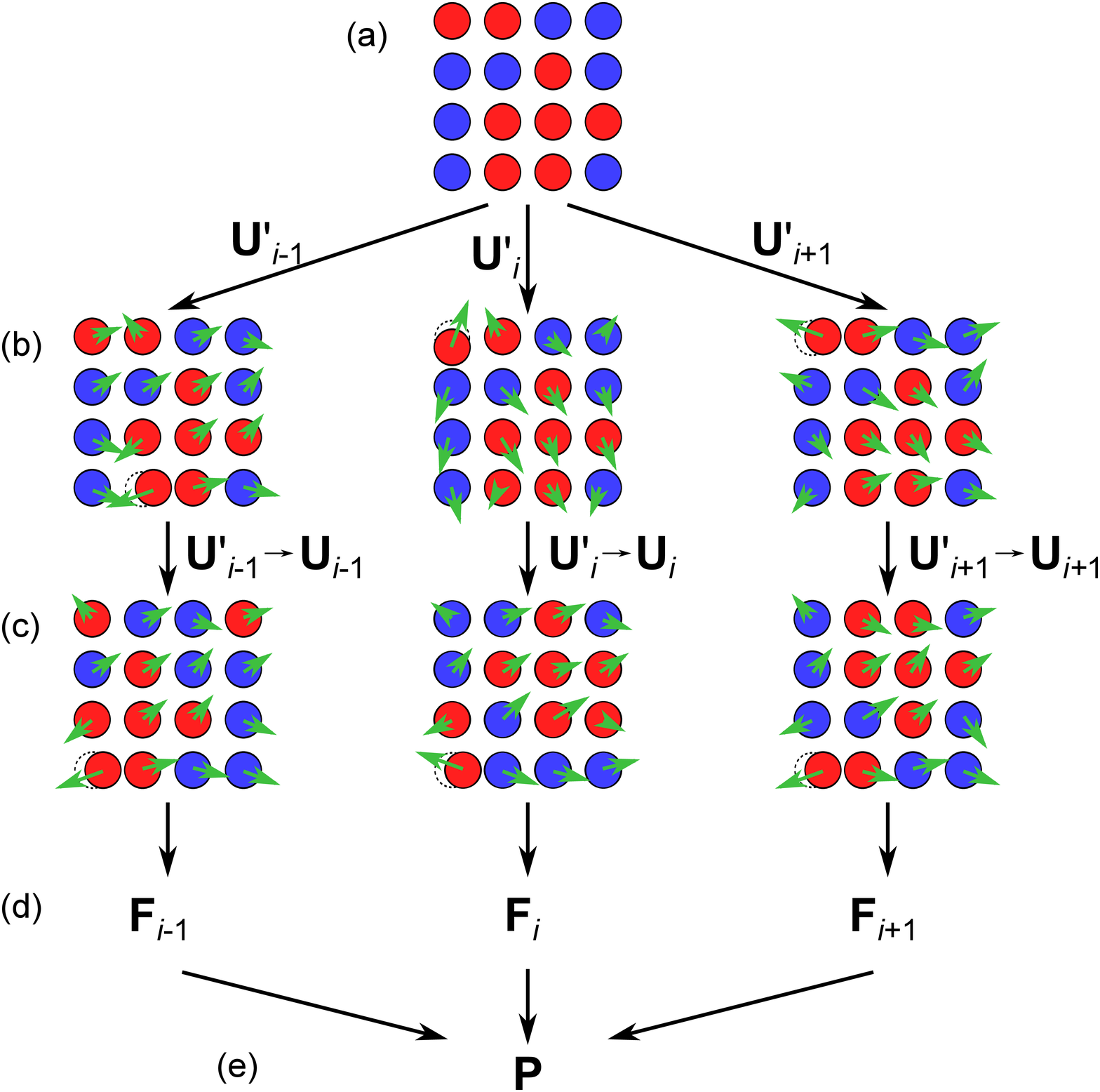}
\end{center}
\caption{
(Color online)
Two dimensional schematic illustration of the procedure to obtain the force constants
of the PM phase.
Red and blue circles represent atoms with spin-up and spin-down magnetic moments,
respectively.
Green small arrows on the circles represent the forces acting on the atoms.
(a) The completely disordered spin configuration is mimicked by the binary SQS.
(b) Symmetrically inequivalent displacements $\mathbf{U}'_i$ are applied to the SQS.
(c) The displaced atoms are moved to one of the crystallographically equivalent sites
of the supercell without magnetic moments.
By this operation, $\mathbf{U}'_i$ is changed to $\mathbf{U}_i$.
(d) The forces acting on the atoms $\mathbf{F}_{i}$ are calculated.
(e) From all the configurations with the displacements, the forces are collected,
and the force constants of the PM phase are obtained via Eq.~(\ref{eq:force_constants}).
\label{fig:symmetrization}
}
\end{figure}
To calculate the force constants of the PM phase,
a symmetrization procedure is applied as follows:
First, we consider the binary SQS to mimic a completely disordered spin configurations
(Fig.~\ref{fig:symmetrization}(a)).
For the SQS, symmetrically inequivalent displacements
$\mathbf{U}'_i (R_{1'} \tau_{1'})$
are applied
(Fig.~\ref{fig:symmetrization}(b)).
%Due to the disordered local magnetic moments,
%the number of the displacements is generally more than that for the FM and NM
%supercells described above.
Next, for the system with
$\mathbf{U}'_i (R_{1'} \tau_{1'})$,
we apply a symmetry operation which moves the displaced atom $R_{1'} \tau_{1'}$
to $R_1 \tau_1$,
where $R_1 \tau_1$ is the atom on the crystallographically equivalent site to
that for the atom $R_{1'} \tau_{1'}$ 
when we neglect the local magnetic moments.
By this operation, the displacement $\mathbf{U}'_i (R_{1'} \tau_{1'})$ is
changed to $\mathbf{U}_i (R_1 \tau_1)$
(Fig.~\ref{fig:symmetrization}(c)).
For the system with $\mathbf{U}_i(R_1\tau_1)$,
a force acting on the atom $R_2\tau_2$, $\mathbf{F}_i (R_2\tau_2)$, is obtained
(Fig.~\ref{fig:symmetrization}(d)).
From all systems, the displacements $\mathbf{U}_i (R_1\tau_1)$
and the forces $\mathbf{F}_i(R_2\tau_2)$ are collected,
and the force constants $\mathbf{P} (R_1\tau_1, R_2\tau_2)$ are calculated
via Eq.~(\ref{eq:force_constants}) as well as the FM and NM cases
(Fig.~\ref{fig:symmetrization}(e)).

In this study, we assume that
many completely disordered spin configurations appear
in a shorter time than that for the thermal atomic fluctuations in the PM phase.
The present averaging process includes effects from many
disordered spin configurations,
and hence we can represent the PM phase by this process.
By considering the site symmetry around the displaced atom,
the symmetry of the original structure can be recovered by the same procedure
as the FM and NM cases.

The force constants obtained by this procedure are regarded as those of the PM phase.
For example, if we use the SQS constructed from the $2 \times 2 \times 2$ supercell
of the conventional bcc unit cell, the obtained force constants can be regarded as
those for the $2 \times 2 \times 2$ supercell in the PM phase.
These PM force constants have the same symmetry as those of
the supercell without magnetic moments.
This approach can be applied not only to the disordered magnetic states but also
to chemically disordered compounds.
Actually, the present procedure is equivalent to that introduced
in Ref.~\citenum{Kormann2012}.
The only difference is that in the present case, we first collect forces on
the atoms, while for the procedure in Ref.~\citenum{Kormann2012},
they first made the force constants from each spin configuration and then
took average of the set of the force constants.
We found that both the procedures provide almost the same result
for the PM bcc Fe.

The calculations of phonon frequencies
%and the mode Gr\"uneisen parameter
are performed by the PHONOPY code.
\cite{Togo2008} % A. Togo et al., Phys. Rev. B 78, 134106 (2008)

%%%%%%%%%%%%%%%%%%%%%%%%%%%%%%%%%%%%%%%%
\subsection{Conditions for electronic structure calculations}
%%%%%%%%%%%%%%%%%%%%%%%%%%%%%%%%%%%%%%%%

For the first-principles electronic structure calculations,
the plane-wave basis projector augmented wave (PAW) method
\cite{Blochl1994} % P. E. Bl\”{o}chl, Phys. Rev. B 50, 17953 (1994)
is employed in the framework of
%density-functional theory
DFT
within the generalized gradient approximation
% (GGA)
in the Perdew-Burke-Ernzerhof form
\cite{Perdew1996} % J. P. Perdew, K. Burke, and M. Ernzerhof, Phys. Rev. Lett. 77, 3865 (1996)
as implemented in the VASP code.
\cite{
Kresse1995, % G. Kresse, J. non-Cryst. Solids 192-193, 222 (1995),
Kresse1996, % G. Kresse and J. Furthmüller, Comput. Mater. Sci. 6, 15 (1996),
Kresse1999} % G. Kresse and D. Joubert, Phys. Rev. B 59, 1758 (1999),
A plane-wave energy cutoff of 300~eV is used.
The radial cutoff of the PAW potential of Fe is 1.22~\AA.
The $3d$ and $4s$ electrons for Fe are treated
as valence and the remaining electrons are kept frozen.
The Brillouin zones are sampled by a $\Gamma$-centered $16\times16\times16$ $k$-point mesh
per conventional bcc unit cell,
and the Methfessel-Paxton scheme~\cite{Methfessel1989} with a smearing width of 0.1 eV
is employed.
%The Brillouin zones of the conventional unit cells are sampled by 16 $\times$ 16 $\times$ 16.
%\textit{k}-point meshes generated in accordance with the $\Gamma$-centered scheme.
The total energy is minimized until the energy convergence becomes less than
1 $\times$ 10$^{-8}$ eV.
%To obtain the equilibrium structures of unit cells at applied pressures,
%internal atomic positions were optimized until the resudual forces became less than
%1 \times 10$^{-3}$ eV / \AA.
%Lattice parameters were optimized
%until the difference between each compoment of the specifiedand 
%and calculated stress became less than 1 \times 10$^{-2}$ GPa.
%
% added on 19/03/2014
%
For the calculations of the PM phase,
the difference between the numbers of spin-up and spin-down electrons is fixed
to be zero.

%%%%%%%%%%%%%%%%%%%%%%%%%%%%%%%%%%%%%%%%
\section{RESULTS AND DISCUSSION}
\label{sec:results}
%%%%%%%%%%%%%%%%%%%%%%%%%%%%%%%%%%%%%%%%
\subsection{Dependence on the selection of SQS}
%%%%%%%%%%%%%%%%%%%%%%%%%%%%%%%%%%%%%%%%

\begin{figure}[tbp]
\begin{center}
\includegraphics[width=\linewidth]{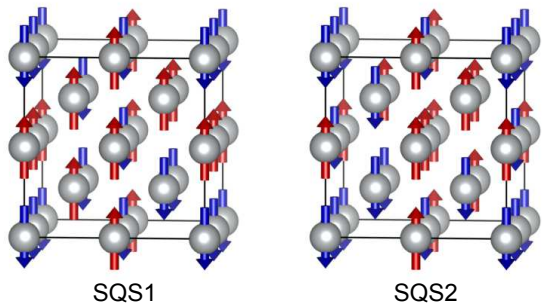}
\end{center}
\caption{
(Color online)
Two kinds of binary SQSs for the bcc structure investigated in this study.
These SQSs are constructed from a supercell obtained by $2 \times 2 \times 2$
isotropic expansion of the conventional bcc unit cell and hence include 16 atoms.
The completely disordered collinear spin configuration is mimicked by these SQSs.
Up (red) and down (blue) arrows represent atoms with spin-up and spin-down
magnetic moments, respectively.
Visualization is performed using the VESTA code.
\cite{Momma2011}
\label{fig:sqs}
}
\end{figure}
\begin{table}[tbp]
\caption{
Pair correlation functions for the SQSs
shown in Fig.~\ref{fig:sqs}.
For SQS1 and SQS2, the pair correlation functions up to the fifth and the fourth
nearest neighbor are optimized by the least absolute deviation method,
respectively.
\label{tb:sqs}
}
\begin{center}
\begin{ruledtabular}
\begin{tabular}{ccc}
Pair correlation function & \multicolumn{1}{c}{SQS1} & \multicolumn{1}{c}{SQS2} \\
\hline
1st NN & $ 0  $ & $ 0$ \\
2nd NN & $ 0  $ & $ 0$ \\
3rd NN & $-1/3$ & $ 0$ \\
4th NN & $ 0  $ & $ 0$ \\
5th NN & $ 0  $ & $-1$ \\
\end{tabular}
\end{ruledtabular}
\end{center}
\end{table}
\begin{table}[tbp]
\begin{center}
\caption{
Calculated equilibrium volume
%with the lowest energy
at the pressure of 0 GPa and the temperature of 0 K
for the FM, PM, and NM bcc Fe.
Energy differences between the magnetic phases are also described.
The energy of the FM phase is set to the origin for the energy differences.
For the PM phase, values derived from the two kinds of SQSs are shown,
and the SQSs for the values are given in the parentheses in the first column.
Note that zero point vibrational contributions are not included.
\label{tb:volume}
}
\begin{ruledtabular}
\begin{tabular}{ccc}
Magnetic phase & Volume (\AA$^3$/atom) & $\Delta E$ (eV/atom) \\
\hline
FM             & 11.3                  & 0.00 \\
PM(SQS1)       & 11.3                  & 0.20 \\
PM(SQS2)       & 11.4                  & 0.19 \\
NM             & 10.5                  & 0.47 \\
%AFM            & 10.9                  & 0.44 \\
\end{tabular}
\end{ruledtabular}
\end{center}
\end{table}
First, we investigate the dependence of the results for the PM phase on the
pair correlation functions of the SQS.
For this purpose,
two kinds of binary SQSs which have different correlation functions are considered.
These two SQSs, i.e., SQS1 and SQS2, are described in Fig.~\ref{fig:sqs}.
Table~\ref{tb:sqs} shows the values of their pair correlation functions.
The pair correlation functions up to the fifth and the fourth nearest neighbor
are optimized for SQS1 and SQS2 by the least absolute deviation method,
respectively.
Table~\ref{tb:volume} shows the calculated equilibrium volume
at the pressure of 0~GPa and the temperature of 0~K, $V_0$,
for the FM, PM, and NM bcc Fe.
For the PM phase, we find that the two SQSs provide almost the same volumes and energies.
%This implies that there is little dependence of properties of the PM states
%on the SQS configurations.
%Actually, we confirmed that all the analyses in the present study showed the same behavior.
This indicates that the results of the PM phase do not depend on the details
of the correlation functions of the SQSs.
In other words, if the values of several principal pair correlation functions
are chosen to be equal to those of the completely disordered spin configuration,
the difference of the results is negligible.
Therefore, we will hereafter show only the results from the SQS1
for the PM phase.

\subsection{Phonon dispersion relations}

\begin{figure*}[tbp]
\begin{center}
\includegraphics[width=\linewidth]{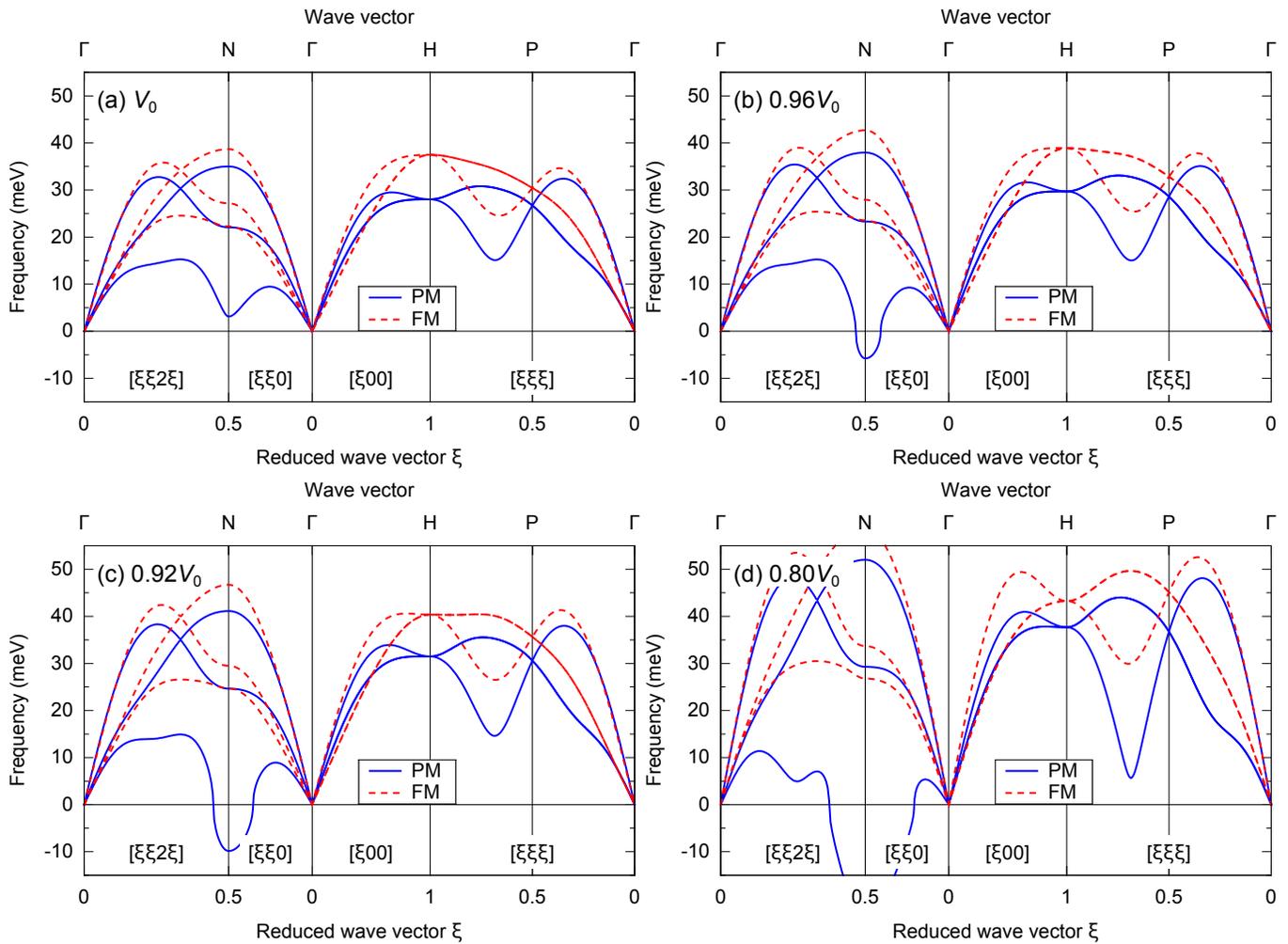}
\end{center}
\caption{
(Color online)
Calculated phonon dispersion relations at the volume of
(a) $V_0$, (b) $0.96 V_0$,
(c) $0.92 V_0$, and (d) $0.80 V_0$
for the bcc Fe, where $V_0$ is the equilibrium volume
of each magnetic phase shown in Table~\ref{tb:volume}.
Blue solid and red dashed curves are for the PM and FM phases,
respectively.
Note that a negative value of a phonon frequency corresponds to an imaginary mode.
\label{fig:phonon}
}
\end{figure*}
%
%Next, volume dependence of the phonon frequency is analyzed.
Figure~\ref{fig:phonon} shows calculated phonon dispersion relations
at four values of volumes around the equilibrium one for each magnetic phase,
$V_0$ (shown in Table~\ref{tb:volume}).
It is found that the phonon frequencies of the PM phase tend to be smaller than
those of the FM phase almost in the whole region.
It is also confirmed that the lowest-frequency modes at the N and 2/3[111] points
of the PM phase decrease their frequencies with decrease of volume.
For the FM phase,
such softenings under volume compression cannot be observed.
This indicates that the two phonon modes characterize the structural stability
of the PM bcc Fe.
Hence we will hereafter discuss the behavior of the two phonons
under volume compression in more detail.

At the N point, one of the transverse modes at the N point becomes
softer under volume compression.
(This mode has the irreducible representation of N$_4^-$,
and hence we will hereafter refer to this mode as the N$_4^-$ mode.) 
This mode finally becomes imaginary under a large volume compression.
This means that the PM bcc Fe dynamically unstable under certain pressures
within the harmonic approximation.
From the calculations with 1\% intervals of volume compression,
we find that this imaginary mode appears from 2\% volume compression.

Atomic displacements along the N$_4^-$ mode can be associated with
the well-known Burgers pathway.
\cite{Burgers1934} % [W. G. Burgers, Physica 1, 561 (1934)].                   
This pathway is composed on two types of transformations:
the displacement of adjacent $(110)$ planes in opposite $[1\bar{1}0]$ directions
and the shear deformation along the $[001]$ direction.
The former part corresponds to the N$_4^-$ mode.
The softening of the N$_4^-$ mode
is experimentally observed for several NM metals
such as Ti, Zr, and Hf,
%, and barium.
\cite{
Petry1991, % W. Petry et al., PRB 43, 10933 (1991)}                                 
Heiming1991, % A. Heiming et al., PRB 43, 10948 (1991)},                            
Trampenau1991}
%Takemura1994} % J. Trampenau et al., PRB 43, 10963 (1991)}.                        
%Chen1988} % Y. Chen et al., PRB 37, 283 (1988)                                     
and these metals actually show bcc-hcp phase transitions.
Therefore, the softening of the N$_4^-$ mode has been considered
as a precursor of the phase transitions.

%\subsection{Potential energy curves along the N$_4^-$ mode}

\begin{figure}[tbp]
\begin{center}
\includegraphics[width=\linewidth]{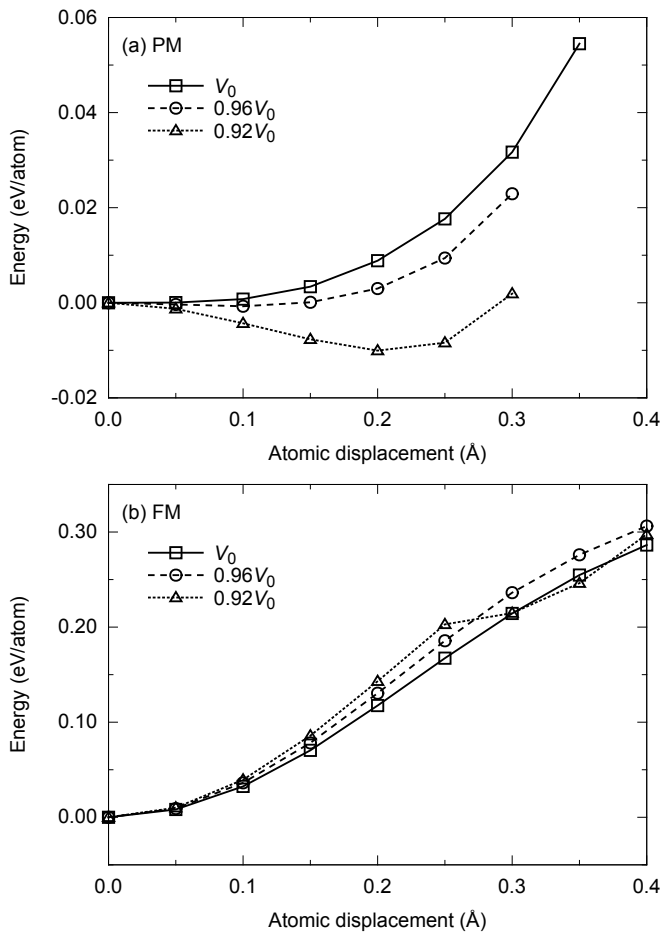}
\end{center}
\caption{
%(Color online)
Potential energy curves along the N$_4^-$ mode
with respect to volumes
for the (a) PM and (b) FM phases.
%at the volume of
%(a) $V_{0,\textrm{FM}}$,
%(b) $0.96V_{0,\textrm{FM}}$,
%and (c) $0.92V_{0,\textrm{FM}}$,
%where $V_{0,\textrm{FM}} = 11.3$\AA$^3$/atom is the equilibrium volume
%for the FM phase.
Square, circle, and triangle symbols are for the results at
$V_0$, $0.96 V_0$, and $0.92 V_0$, respectively,
where $V_0$ denotes the equilibrium volume.
%Insets are the magnified views
%around the energies for the PM phase.
%\textcolor{red}{
%For comparison,
%the result at $0.96 V_0$ obtained by the CPA-DLM method
%\cite{Mankovsky2013}
%is also shown by the green closed triangle symbols.
%}
%for the FM, PM, and NM phases, respectively.
Lines connecting the symbols are guides for the eyes.
%Vertical dotted lines represent the displacement corresponding
%the distorted hcp structures of the Burgers pathway,
%$\sqrt{2} a / 12$, where $a$ is the lattice constant of
%the bcc structure.
\label{fig:pes}
}
\end{figure}
Experimentally, the PM bcc Fe undergoes pressure-induced phase transition
to the fcc phase rather than the hcp one.                                       
However, the computational results for the N$_4^-$ mode in the PM phase clearly
becomes imaginary under volume compression,
which seemingly indicates the spontaneous bcc-hcp phase transition
along the Burgers pathway.
This seeming contradiction can be ascribed to the neglect of
anharmonic effects of the phonons at high temperature.
To analyze anharmonic effects of the N$_4^-$ phonon mode in more detail,
potential energy curves along this mode are calculated around the equilibrium volume,
$V_{0} = 11.3 \textrm{\AA$^3$/atom}$.
The result is shown in Fig.~\ref{fig:pes}.
The energy of the PM phase with the displacements of the atoms is determined
as follows:
First we consider the configurations which are derived from the SQS by applying
symmetry operations of the supercell corresponding to the SQS.
Then we give the atomic displacements along the N$_4^-$ mode to these
configurations.
Finally we take an average of the energies obtained from them,
which is regarded as the energy of the PM phase. 
Note that for the PM phase, several calculations with large atomic displacements
encountered convergence problems.
Therefore, results for several points in this region are not shown for the PM phase.
It is clearly confirmed that the dependence of the energies
on the phonon amplitude is much weaker for the PM phase
than that for the FM phase.
This causes atomic large thermal fluctuations of the atoms for the PM phase.
We can also find that for the PM phase,
the curve gets close to a double-well shape
with decrease of volume.
This means that the anharmonicity along the N$_4^-$ mode becomes stronger
as the volume decreases.
This anharmonicity indicates that
time-averaged atomic positions become the ideal ones for the bcc structure
due to the large thermal fluctuations
at sufficiently high temperature.
Therefore, the bcc structure is stabilized along the N$_4^-$ mode
at the temperature and does not causes the spontaneous transition to the hcp phase.
This idea is well consistent with experimental results.

\begin{table*}[tbp]
\begin{center}
\caption{
\label{tb:fitting}
Fitting parameters to the calculated potential energy curves.
The fourth-order polynomial functions $E(x) = E_0 + {a_2 x^2/ 2} + {a_4 x^4/ 4!}$
is considered,
where $x$, $E(x)$, $E_0$, $a_2$, and $a_4$ are the atomic displacement,
the energy at the corresponding displacements,
the energy at the ideal bcc positions,
the coefficient for the second order, and
the coefficient for the fourth order, respectively.
The coefficients $a_2$ and $a_4$ are the fitting parameters
and determined by means of the least-squares method.
%The interval of the displacements for the fitting is set to 0.05\AA.
}
\begin{ruledtabular}
\begin{tabular}{ccddd}
\textrm{Volume} &
\textrm{Magnetic phase} &
\multicolumn{1}{c}{$a_2$ (eV/(atom$\cdot$\AA$^2$))} &
\multicolumn{1}{c}{$a_4$ (eV/(atom$\cdot$\AA$^4$))} &
\multicolumn{1}{c}{$a_4/a_2$ (\AA$^{-2}$)}\\
\hline
$V_{\textrm{0}}$
 & \textrm{PM} & 0.2  & 66.9  & 324.1  \\
 & \textrm{FM} & 6.4  & -217.3  & -33.7  \\
% & \textrm{PM2} & 0.4  & 61.6  & 172.2  \\
% & \textrm{NM} & -9.0  & 563.3  & -62.3  \\
\hline
$0.96V_{\textrm{0}}$
 & \textrm{PM} & -0.2  & 89.2  & -558.9  \\
 & \textrm{FM} & 7.2  & -257.7  & -35.7  \\
% & \textrm{PM2} & -0.4  & 111.9  & -298.3  \\
% & \textrm{NM} & -9.6  & 614.7  & -64.0  \\
\hline
$0.92V_{\textrm{0}}$
 & \textrm{PM} & -0.9  & 131.9  & -139.1  \\
 & \textrm{FM} & 7.3  & -287.0  & -39.1  \\
% & \textrm{PM2} & -1.6  & 321.9  & -207.6  \\
% & \textrm{NM} & -10.2  & 673.1  & -65.9  \\
\end{tabular}
\end{ruledtabular}
\end{center}
\end{table*}
To evaluate the anharmonicity of these curves more quantitatively,
the curves are fitted by the fourth-order polynomial functions,
$E(x) = E_0 + {a_2 x^2/ 2} + {a_4 x^4/ 4!}$,
where $x$, $E(x)$, $E_0$, $a_2$, and $a_4$ are the displacements of atoms,
the energy at the corresponding displacements,
the energy at the ideal bcc positions,
the coefficient for the second order, and
the coefficient for the fourth order, respectively.
The fitting parameters, $a_2$ and $a_4$, are determined
by means of the least-squares method.
%The interval of the displacements for the fitting is set to 0.05\AA.
The result is shown in Table~\ref{tb:fitting}.
The ratio between the second- and the fourth-order coefficients, $a_4/a_2$,
are also shown in this table.
This ratio clarifies the fourth-order effect with respect to the second-order
harmonic effect and can be considered as one of the indices to evaluate
the strength of the anharmonicity.
It is shown that the absolute values of the ratios for the PM phase
are much larger than that for the FM phase.
This reveals strong anharmonicity of the N$_4^-$ mode for the PM phase.
%For the FM state, the second-order coefficient becomes larger as the volume
%decreases, and the fourth-order coefficient shows the opposite behavior.
%For the PM state, on the other hand,
%the second-order coefficient becomes smaller with decreasing volume,
%and the fourth-order coefficient has the opposite trend.
%%The behavior of the second-order coefficients corresponds to the harmonic phonon
%%frequencies shown in Fig.~\ref{fig:squared_frequency}.
We can also find that $a_4$ of the PM phase is positive at all of
the three volumes,
which makes the potential energy curve close to the double-well shape.
%This means that time-averaged positions of the atoms become the ideal positions
%of the bcc structure at sufficiently high temperature.
%Therefore, the bcc structure is stabilized along the N$_4^-$ mode
%at sufficiently high temperature.

%Recently, Mankovsky \textit{et al.} have reported
Recently, \citeauthor{Mankovsky2013} have reported
\cite{Mankovsky2013}
the potential energy curves
along the N$_4^-$ mode of the PM bcc Fe based on
the coherent potential approximation (CPA)
in combination with the disordered local moment (DLM) scheme.
%Note that The inconsistency between the two methods is probably caused by the
%difference of the treatment for
%the microscopic atomic structure;
%in the CPA-DLM method,
%both spin-up and spin-down atoms in disordered spin configurations
%are represented by an effective potential,
%while in the present SQS-based approach,
%the disordered spin-up and spin-down atoms are considered explicitly.
Their result at the volume of 10.4\AA/atom ($\approx 0.92 V_0$) showed that
the energy of the PM bcc Fe along the N$_4^-$ mode decreased monotonically
(see Fig.~4(c) in Ref.~\citenum{Mankovsky2013}).
This result is different from that from the present SQS-based approach which
shows the double-well potential.
The major source of the inconsistency may be caused by the anharmonic effects of
the phonons rather than the difference of the theoretical treatment for the
PM phases.

In Fig.~\ref{fig:phonon}, we can also find the softening of
the longitudinal phonon mode at the 2/3[111] point under volume compression.
This mode consists in the displacement of the $(111)$ planes along the $[111]$ direction,
which corresponds to a bcc-$\omega$ phase transition pathway.
Similar to the case of the phonon mode at the N point,
the softening has been experimentally observed for the NM metals
such as Ti, Zr, and Hf,
which actually show the bcc-$\omega$ phase transitions.                                 
\cite{
Petry1991, % W. Petry et al., PRB 43, 10933 (1991)}                                 
Heiming1991, % A. Heiming et al., PRB 43, 10948 (1991)},                            
Trampenau1991}                                                                      
Recently, our group proposed
\cite{Togo2013}                                                                
a novel bcc-fcc phase transition pathway
where the $\omega$ structure appears as an energy barrier
during the bcc-fcc transformation.
The present finding implies that the softening of the longitudinal mode
at the 2/3[111] point can be considered as a precursor for the bcc-fcc transition
as well as that for the bcc-$\omega$ transition.
Actually, this is consistent with the experimental fact that
the PM bcc Fe undergoes the phase transition to the PM fcc Fe
under pressure.

%%%%%%%%%%%%%%%%%%%%%%%%%%%%%%%%%%%%%%%%
\section{SUMMARY}
%%%%%%%%%%%%%%%%%%%%%%%%%%%%%%%%%%%%%%%%

%With special interests in the pressure-induced phase transition,
Structural stability of the PM bcc Fe under pressure is analyzed in detail
based on the phonon dispersion relations.
The PM phase is described using the binary SQS.
In order to obtain the phonon frequencies for the PM phase,
symmetrization procedure is applied.
For the PM phase,
the low-frequency transverse mode at the N point,
which correspond to the bcc-hcp phase transition pathway,
shows strong softening under volume compression.
This mode becomes imaginary by 2\% volume decrease
within the harmonic approximation.
This result seems contradictory to the experimental fact that
phase transition from the PM bcc to hcp phases does not occur under compression.
However, this puzzle can be solved by taking anharmonic behavior of
the N$_4^-$ into consideration;
potential energy curves along the N$_4^-$ mode for the PM phase is found
to become a double-well shape under compression.
%This clear anharmonicity indicates that the PM bcc iron becomes stable
%along this mode at sufficiently high temperature,
%and hence the PM bcc iron at such temperature does not undergo
%the spontaneous transition to the hcp phase under pressure.
%This result corresponds to experimental reports.
%
We also find the softening at the longitudinal phonon mode at the 2/3[111] point
under volume compression.
This mode can be associated with the bcc-fcc phase transition pathway
where the $\omega$ structure appears during the transformation.
Hence the softening of this mode can be considered as
a precursor of the phase transition from PM bcc to PM fcc phases,
which is indeed observed experimentally.
Such softening behaviors are not observed for the FM phase,
which implies that the phase transition mechanism is essentially different
between the FM and PM phases.

%%%%%%%%%%%%%%%%%%%%%%%%%%%%%%%%%%%%%%%%

\begin{acknowledgments}

The authors thank F. K\"{o}rmann for valuable discussions.
This work was supported by MEXT Japan through Elements Strategy Initiative for
Structural Materials (ESISM) of Kyoto University.

\end{acknowledgments}

%\bibliography{main.bib}

\end{document}